\documentclass[11pt,a4paper]{article}

\newif\ifanonymized
\anonymizedfalse  % <-- Change to \anonymizedtrue for double-blind submission

\usepackage[utf8]{inputenc}
\usepackage[T1]{fontenc}
\usepackage{lmodern}
\usepackage[english]{babel}
\usepackage{geometry}
\usepackage{graphicx}
\usepackage{booktabs}
\usepackage{hyperref}
\usepackage{amsmath}
\usepackage{cleveref}
\usepackage{enumitem}
\usepackage{xcolor}
\usepackage{listings}
\usepackage{authblk}
\usepackage{microtype}
\usepackage{tabularx}
\usepackage{longtable}

\hypersetup{
  colorlinks=true,
  linkcolor=blue!70!black,
  citecolor=blue!70!black,
  urlcolor=blue!70!black,
}

\title{Less Is More: Engineering Challenges of On-Device Small Language Model Integration in a Mobile Application}

\ifanonymized
  \author{[Anonymous for review]}
  \date{}
\else
  \author{William Oliveira\\
  Independent Researcher\\
  \texttt{contact@woliveiras.com}}
  \date{April 2026}
\fi

\begin{document}

% --- Title page ---
\begin{titlepage}
  \centering
  \vspace*{\fill}
  {\LARGE\bfseries Less Is More: Engineering Challenges of On-Device Small Language Model Integration in a Mobile Application\par}
  \vspace{2em}
  \ifanonymized
    {\large [Anonymous for review]\par}
  \else
    {\large William Oliveira\par}
    \vspace{0.5em}
    {\normalsize Independent Researcher\par}
    {\normalsize \texttt{contact@woliveiras.com}\par}
    \vspace{1em}
    {\normalsize April 2026\par}
  \fi
  \vspace*{\fill}
\end{titlepage}

% --- Abstract page ---
\begin{titlepage}
  \vspace*{\fill}
  \begin{abstract}
On-device Small Language Models (SLMs) promise fully offline, private AI experiences for mobile users (no cloud dependency, no data leaving the device). But is this promise achievable in practice? This paper presents a longitudinal practitioner case study documenting the engineering challenges of integrating SLMs (Gemma 4 E2B, 2.6B parameters; Qwen3 0.6B, 600M parameters) into Palabrita, a production Android word-guessing game. Over a 5-day development sprint comprising 204 commits ($\sim$90 directly AI-related), the system underwent a radical transformation: from an ambitious design where the LLM generated complete structured puzzles (word, category, difficulty, and five hints as JSON) to a pragmatic architecture where curated word lists provide the words and the LLM generates only three short hints, with a deterministic fallback if it fails. We identify five categories of failures specific to on-device SLM integration: output format violations, constraint violations, context quality degradation, latency incompatibility, and model selection instability. For each failure category, we document the observed symptoms, root causes, and the prompt engineering and architectural strategies that effectively mitigated them, including multi-layer defensive parsing, contextual retry with failure feedback, session rotation, progressive prompt hardening, and systematic responsibility reduction. Our findings demonstrate that on-device SLMs are viable for production mobile applications, but only when the developer accepts a fundamental constraint: the most reliable on-device LLM feature is one where the LLM does the least. We distill our experience into eight actionable design heuristics for practitioners integrating SLMs into mobile apps.
  \end{abstract}
  \vspace*{\fill}
\end{titlepage}

% --- Table of contents ---
\tableofcontents
\thispagestyle{empty}
\newpage

% ============================================================
\section{Introduction}
\label{sec:introduction}
% ============================================================

The deployment of Large Language Models on mobile devices has emerged as a compelling alternative to cloud-based inference. On-device execution offers zero-latency privacy guarantees (user data never leaves the device), eliminates network dependency, and removes per-query cost. Frameworks such as Google's LiteRT-LM~\cite{litert}, MLC LLM~\cite{mlcllm}, and Android's AICore~\cite{aicore} have made it technically possible to run models with hundreds of millions to billions of parameters directly on smartphones. Google's AI Edge Gallery showcase app~\cite{litert} and the growing catalog of \texttt{.litertlm} models on Hugging Face demonstrate that the infrastructure is maturing rapidly.

But a significant gap exists between infrastructure readiness and engineering guidance. While abundant research benchmarks SLM performance on standardized tasks (measuring tokens per second, perplexity, and task accuracy), almost nothing documents the \emph{engineering reality} of shipping an SLM-powered feature in a production mobile application. What happens when a 600MB model consistently wraps its JSON output in markdown code fences? When it generates words with the wrong number of letters despite explicit instructions? When inference sessions degrade in quality after three generations due to KV cache saturation? These are not hypothetical scenarios; they are the daily challenges a practitioner faces, and the solutions require a combination of prompt engineering, defensive parsing, and architectural pragmatism that existing literature does not address.

This paper presents a practitioner case study of integrating on-device SLMs into Palabrita, a Wordle-style word-guessing game for Android. The study is based on the complete Git history of the project: 204 commits over 5 days, with $\sim$90 commits directly related to AI/LLM integration. The commit log serves as a contemporaneous, tamper-evident record of every decision, failure, and course correction, providing an unusually rigorous evidence trail for a case study.

The story of this integration is one of progressive disillusionment followed by pragmatic adaptation. The initial design was ambitious: the LLM would generate complete puzzles (a word, its category, a difficulty rating, a rarity indicator, and five progressive hints) as structured JSON. By the end of five days, the LLM's sole responsibility was writing three short hints about a word selected from a curated list, with a deterministic fallback if it failed. This trajectory was not planned; it was \emph{discovered} through repeated contact with the constraints of on-device inference.

We investigate three research questions:

\begin{itemize}
  \item \textbf{RQ1:} Is it viable today to use on-device LLMs in Android applications to deliver a fully offline, private user experience?
  \item \textbf{RQ2:} What types of failures emerge when running LLM inference locally on Android without sending data to the cloud?
  \item \textbf{RQ3:} What prompt engineering and architectural strategies effectively mitigate these failures?
\end{itemize}

Our contributions are: (1) a taxonomy of five failure categories specific to on-device SLM integration, documented with commit-level evidence and code examples; (2) the observation of a \emph{progressive simplification arc} where the LLM's responsibility was systematically reduced from seven output fields to zero for word generation, converging on hint-only generation as the reliable equilibrium; (3) a detailed account of the prompt engineering iteration process, showing four complete rewrites of the prompt system with before-and-after comparisons; (4) a case study of an architectural anti-pattern (using Android's WorkManager for foreground LLM operations) and its resolution; and (5) eight actionable design heuristics derived from these observations.

The remainder of this paper is organized as follows. \Cref{sec:background} reviews the relevant background on SLMs, inference frameworks, and prompt engineering. \Cref{sec:study-design} describes the study methodology, subject system, and data collection approach. \Cref{sec:timeline} presents a day-by-day development timeline that provides narrative context for the technical findings. \Cref{sec:rq2-rq3-findings} presents the failure taxonomy and progressive simplification arc. \Cref{sec:strategies} details the mitigation strategies with code examples. \Cref{sec:workmanager} presents the WorkManager case study. \Cref{sec:discussion} discusses design heuristics and implications. \Cref{sec:threats} addresses threats to validity, and \cref{sec:conclusion} concludes.

% ============================================================
\section{Background and Related Work}
\label{sec:background}
% ============================================================

\subsection{Small Language Models}

Small Language Models (SLMs), typically models with fewer than 3 billion parameters, have become the practical choice for on-device deployment. The Gemma family (Google DeepMind) offers models from 1B to 4B parameters in quantized formats optimized for edge devices~\cite{gemma}. Qwen3 (Alibaba) provides models starting at 0.6B parameters~\cite{qwen}. Phi-4 Mini (Microsoft) targets the 3.8B parameter range~\cite{phi4}. These models sacrifice capability compared to cloud models (GPT-4, Claude) but gain the ability to run entirely on consumer hardware.

The capability gap between cloud and on-device models is not merely quantitative (fewer tokens per second, slightly worse accuracy). It is \emph{qualitative}: SLMs exhibit different failure modes that require different engineering approaches. A cloud model prompted to output JSON will almost always produce valid JSON. An SLM prompted identically may wrap its JSON in markdown code fences, translate JSON keys into the output language, truncate the response mid-token, or produce valid JSON with semantically wrong values. These qualitative differences are poorly documented in the literature, which tends to evaluate SLMs on standardized benchmarks rather than on the constrained, application-specific tasks that practitioners need.

For our use case (generating short game hints about known words), the key question is not whether SLMs match cloud models on benchmarks, but whether they can reliably produce simple structured output under tight constraints. Our findings show this reliability is significantly lower than cloud models suggest, requiring extensive engineering compensations.

\subsection{On-Device Inference Framework Evaluation}
\label{sec:framework-eval}

Before implementation, we evaluated three inference frameworks for Android. This evaluation is itself a contribution, as no published comparison exists for this specific decision space. \Cref{tab:framework-comparison} summarizes the key dimensions.

\begin{table}[ht]
\centering
\small
\caption{Comparison of on-device inference frameworks for Android.}
\label{tab:framework-comparison}
\begin{tabularx}{\textwidth}{lXXX}
\toprule
\textbf{Dimension} & \textbf{LiteRT-LM} & \textbf{MLC LLM} & \textbf{AICore} \\
\midrule
Provider & Google DeepMind & CMU (community) & Google Android \\
Prompt control & Full (system + user) & Full & Task APIs + Prompt API (beta, no sampler config) \\
Model choice & Gemma, Qwen, Phi, Llama & Broad (via TVM) & Gemini Nano only \\
Sampler config & temperature, topK, topP & Yes & No \\
Integration effort & 1 Gradle dependency & Rust + NDK + TVM & 1 Gradle dependency \\
Model format & \texttt{.litertlm} (pre-converted) & TVM-compiled & OS-managed \\
Backend selection & GPU / NPU / CPU & GPU / CPU & System-managed \\
Emulator support & Yes (CPU) & No (documented) & No \\
Device support & Any (with CPU fallback) & Any (GPU preferred) & Selected flagships (multi-vendor) \\
Model swapping & Config change only & Re-compile per model & Not possible \\
\bottomrule
\end{tabularx}
\end{table}

\textbf{LiteRT-LM} (Google DeepMind)~\cite{litert} is a production-ready, cross-platform inference framework. It provides a native Kotlin SDK with coroutine support, GPU/NPU/CPU backend selection, Play Asset Delivery integration for model distribution, and full control over prompts and sampler configuration (\texttt{temperature}, \texttt{topK}, \texttt{topP}). It supports Gemma, Qwen, Phi, and Llama model families via pre-converted \texttt{.litertlm} format. Integration requires a single Gradle dependency. The SDK exposes both single-turn inference (\texttt{generateSingleTurn(systemPrompt, userPrompt)}) and multi-turn chat sessions (\texttt{createChatSession(systemPrompt)}) with streaming support.

\textbf{MLC LLM}~\cite{mlcllm} is a community-maintained compiler-based framework originating from CMU research. It offers broad model support via TVM compilation but requires a heavy build toolchain: Rust (for tokenizer cross-compilation), Android NDK, TVM compiler, and per-model compilation steps. Its Android documentation explicitly notes that emulator testing is unsupported and documents a known issue on Adreno hardware: models with weight layouts using the \texttt{\_1} suffix trigger a 20--50 second system UI freeze at the prefill stage (first inference initialization); the \texttt{\_0} suffix layout avoids the issue~\cite{mlcllmandroid}. The model preparation pipeline (clone, compile, package) adds significant friction compared to downloading a pre-converted weight file. For a solo developer on a rapid iteration cycle, this toolchain complexity was a dealbreaker.

\textbf{AICore / ML Kit GenAI}~\cite{aicore} is Android's system-level foundation model service, providing Gemini Nano through a high-level API. While it offers seamless deployment (model managed by the OS), it is restricted to Gemini Nano, a single model with no custom alternatives. While ML Kit also offers a Prompt API (beta) for free-form text prompts, it exposes no sampler configuration (temperature, topK, topP), making fine-grained inference control impossible. Device support spans selected flagships across multiple vendors (Google, Samsung, Xiaomi, OnePlus, and others), but remains limited to specific high-end models with no CPU fallback for broader coverage (Table~\ref{tab:aicore-prompt-devices}). For a game requiring fine-grained prompt control and model selection, AICore was unsuitable.

\begin{table}[h]
\centering
\caption{ML Kit GenAI Prompt API supported devices by Gemini Nano version, consulted May~5, 2026~\cite{mlkitpromptdevices}.}
\label{tab:aicore-prompt-devices}
\small
\begin{tabularx}{\textwidth}{llX}
\toprule
\textbf{Nano version} & \textbf{Vendor} & \textbf{Models} \\
\midrule
nano-v2 & Google   & Pixel 9, Pixel 9 Pro, Pixel 9 Pro XL, Pixel 9 Pro Fold \\
        & Honor    & Magic V5, Magic 7, Magic 7 Pro \\
        & iQOO     & iQOO 13 \\
        & Motorola & Razr 60 Ultra, Razr Ultra 2025 \\
        & OnePlus  & OnePlus 13, OnePlus 13s \\
        & OPPO     & Find N5 \\
        & POCO     & F7 Ultra, F8 Pro, F8 Ultra, X7 Pro, X8 Pro \\
        & realme   & GT 7 Pro \\
        & Samsung  & Galaxy Z Fold7, Galaxy Z TriFold \\
        & Xiaomi   & 14T Pro, 15, 15T, 15T Pro, 15 Ultra, 17, 17 Ultra, Pad Mini \\
        & vivo     & X200 FE, T4 Ultra \\
\midrule
nano-v3 & Google   & Pixel 10, Pixel 10 Pro, Pixel 10 Pro XL, Pixel 10 Pro Fold \\
        & Honor    & Magic 8 Pro \\
        & iQOO     & iQOO 15 \\
        & Motorola & Signature \\
        & OnePlus  & OnePlus 15, OnePlus 15R \\
        & OPPO     & Find X9, Find X9 Pro, Find X8, Find X8 Pro, Reno 14 Pro 5G, Reno 15 Pro 5G, Reno 15 Pro Mini 5G, Reno 15 Pro Max 5G \\
        & realme   & GT 7T \\
        & Samsung  & Galaxy S26, Galaxy S26+, Galaxy S26 Ultra \\
        & vivo     & X200T, X200, X200 Pro, X300, X300 Pro \\
\bottomrule
\end{tabularx}
\end{table}

\textbf{Decision:} We chose LiteRT-LM for three reasons: (1)~\emph{full prompt control}, essential for the iterative prompt engineering documented in \cref{sec:strategies}; (2)~\emph{model flexibility}, which allowed us to swap from Gemma 3 1B to Qwen3 0.6B in under 8 hours without changing the framework; and (3)~\emph{minimal integration friction} (a Gradle dependency versus a multi-tool compilation pipeline). This decision paid off in practice: the ability to rapidly iterate on prompts and swap models was critical to the project's success. In a five-day sprint where four complete prompt rewrites occurred, any additional friction in the model-to-inference pipeline would have consumed days.

\subsection{Prompt Engineering for Constrained Models}

Prompt engineering for cloud LLMs is well-documented~\cite{cot,promptpatterns}. Chain-of-thought prompting~\cite{cot}, few-shot examples, and system/user prompt separation are standard techniques. SLMs, though, behave very differently under prompting:

\begin{itemize}
  \item \textbf{Format compliance is unreliable.} Cloud models almost always produce valid JSON when prompted. SLMs frequently add markdown code fences, mix natural language with JSON, or translate JSON keys.
  \item \textbf{Numerical reasoning is weak.} Counting characters in a word (e.g., ``does `estufa' have 5 or 6 letters?'') is unreliable at the sub-3B parameter scale.
  \item \textbf{Language drift is systematic.} English-dominant training data causes SLMs to default to English output even when explicitly instructed otherwise, especially for structured data like JSON keys.
  \item \textbf{Emphasis tokens have measurable impact.} Words like ``CRITICAL'' and ``NEVER'' before constraints improved compliance in our experiments, suggesting SLMs are more sensitive to token-level emphasis than large models.
  \item \textbf{ISO codes are ambiguous.} SLMs interpret ``pt'' unreliably (Portuguese? Part? Point?). Full language names (``Brazilian Portuguese'') are unambiguous.
\end{itemize}

Existing work on structured output from LLMs focuses on function calling and tool use in large models~\cite{toolformer}; we found these approaches ineffective at the SLM scale. Function calling requires the model to understand a schema contract; SLMs at 0.6B parameters lack this capability. Our initial prompt referenced ``the provided function'' (a vestige from a function-calling design that was never implemented), and the model had no idea what function to call.

\subsection{Related Experience Reports}

Experience reports documenting on-device ML deployment exist for classical ML tasks (image classification, speech recognition) via TensorFlow Lite and Core ML~\cite{tflite,coreml}. However, on-device \emph{generative} AI, where the model produces variable-length, unpredictable natural language output, presents fundamentally different challenges. Classical ML tasks produce fixed-size outputs (a class label, a bounding box, a waveform); generative tasks produce unbounded text that must be parsed, validated, and integrated into application state.

The closest related work is Google's AI Edge Gallery~\cite{litert}, an experimental showcase app that demonstrates on-device inference with multiple models. Yet no published engineering analysis of its development process exists, and showcase apps are designed for demonstration rather than production use. They do not face the parsing, validation, and fallback challenges documented here.

Industry blog posts from mobile AI practitioners describe similar challenges anecdotally, but without the systematic evidence (commit-level traceability, failure taxonomy, quantified frequencies) that enables reproducibility and generalization.

% ============================================================
\section{Study Design}
\label{sec:study-design}
% ============================================================

\subsection{Author Profile and Methodology}

This is a practitioner case study conducted by a senior software engineer with expertise in AI Engineering and multi-platform development. The author was the sole developer of the system, providing complete visibility into every design decision and its rationale. The methodology follows Runeson and H\"ost's guidelines for case study research in software engineering~\cite{runeson}, adapted for a single-developer context where the commit log serves as the primary data source.

The development followed a \emph{Spec Driven Development} workflow: specifications were written before code, acceptance criteria became tests, and implementation proceeded until tests passed. This discipline meant that each design decision, including reversals, was documented in specifications before being implemented.

\subsection{Subject System}

Palabrita is a Wordle-style word-guessing game for Android. Players guess a hidden word within six attempts, guided by color-coded feedback (correct letter in correct position, correct letter in wrong position, letter not in word) and natural language hints generated by an on-device LLM. The game supports three languages (Portuguese, English, Spanish) with an extensible architecture for adding more.

The application is built with Kotlin, Jetpack Compose, Hilt for dependency injection, Room for persistence, and a multi-module MVVM architecture with unidirectional data flow. The project comprises 9 Gradle modules:

\begin{itemize}
  \item \texttt{app/} --- Application module (DI, navigation, theme)
  \item \texttt{core/ai/} --- LLM engine, puzzle generation, prompt management, response parsing
  \item \texttt{core/common/} --- State machine, device capabilities, shared utilities
  \item \texttt{core/model/} --- Domain models and repository interfaces
  \item \texttt{core/data/} --- Room database, DAOs, repository implementations
  \item \texttt{feature/game/} --- Game screen, keyboard, game logic
  \item \texttt{feature/home/} --- Home screen with puzzle queue
  \item \texttt{feature/onboarding/} --- Onboarding flow (language, model selection)
  \item \texttt{feature/settings/} --- Settings, statistics, model management
\end{itemize}

Minimum SDK is Android 12 (API 31). The AI subsystem is isolated in the \texttt{core/ai} module, decoupled from the rest of the application through three interfaces: \texttt{LlmEngineManager} (engine lifecycle), \texttt{PuzzleGenerator} (puzzle/hint generation), and \texttt{PromptProvider} (prompt templates). This isolation enabled rapid iteration on the AI subsystem without affecting other modules.

\subsection{Models Under Study}
\label{sec:models}

Over the development period, seven models appeared in the codebase. \Cref{tab:models-full} lists all of them, distinguishing those that were actually downloaded and executed from those that were registered as candidates but never run.

\begin{table}[ht]
\centering
\small
\caption{Complete inventory of models that appeared in the codebase. \emph{Tested} indicates the model was downloaded and executed; \emph{Shipped} indicates it reached the production configuration.}
\label{tab:models-full}
\begin{tabular}{lcccccl}
\toprule
\textbf{Model} & \textbf{Params} & \textbf{Size} & \textbf{RAM} & \textbf{Tested} & \textbf{Shipped} & \textbf{Outcome} \\
\midrule
Gemma 4 E2B      & $\sim$2.6B & 2{,}583\,MB & $\geq$8\,GB  & Yes & Yes & Premium tier \\
Gemma 3 1B       & $\sim$1B   & 1{,}005\,MB & $\geq$4\,GB  & Yes & No  & Replaced by Qwen3 0.6B (Day 2) \\
Qwen3 0.6B       & $\sim$600M & 614\,MB     & $\geq$2\,GB  & Yes & Yes & Compact tier \\
Gemma 4 E4B      & $\sim$4B   & 3{,}650\,MB & $\geq$12\,GB & No  & No  & Never downloaded: too large \\
Phi-4 Mini       & $\sim$3.8B & 3{,}910\,MB & $\geq$8\,GB  & No  & No  & Never downloaded: too large \\
Qwen2.5 1.5B     & $\sim$1.5B & 1{,}600\,MB & $\geq$4\,GB  & No  & No  & Never downloaded: deprioritized \\
DeepSeek R1 1.5B & $\sim$1.5B & 1{,}830\,MB & $\geq$4\,GB  & No  & No  & Never downloaded: deprioritized \\
\bottomrule
\end{tabular}
\end{table}

Only three models were ever downloaded and executed: Gemma 4 E2B, Gemma 3 1B, and Qwen3 0.6B. The remaining four were registered in a Day 5 consolidation pass (commit \texttt{da341c6}) as evaluation candidates and removed the same day before any inference was attempted (commit \texttt{0e56401}). The barrier was practical: model files ranged from 1.6\,GB to 3.9\,GB. At iterative game development pace, waiting 10--30 minutes per download to test a model that might be discarded within hours was not a viable use of time. This is itself an instance of the constraints documented throughout this paper: hardware and network realities shaped the evaluation boundary before a single inference token was generated.

The final production configuration exposes only two models to users:

\begin{itemize}
  \item \textbf{Gemma 4 E2B} ($\sim$2.6B parameters, 2.6\,GB): Premium tier for devices with $\geq$8\,GB RAM. Better instruction following, more coherent hints, but slower initialization ($\sim$10\,s) and larger download.
  \item \textbf{Qwen3 0.6B} ($\sim$600M parameters, 614\,MB): Compact tier for devices with $\geq$2\,GB RAM. Faster initialization ($\sim$5\,s), smaller download, adequate hint quality, but weaker instruction following and more format violations.
\end{itemize}

Performance measurements were collected on a \textbf{Google Pixel 7 Pro} (12\,GB RAM, Tensor G2). A Pixel 10 Pro (Tensor G5) was available but benchmark data from it was not systematically collected within the 5-day development window. \Cref{tab:benchmarks} summarizes the measured performance.

\begin{table}[ht]
\centering
\small
\caption{Benchmark data collected on a Google Pixel 7 Pro (Tensor G2, 12\,GB RAM).}
\label{tab:benchmarks}
\begin{tabular}{llcccc}
\toprule
\textbf{Device} & \textbf{Model} & \textbf{Backend} & \textbf{Prefill} & \textbf{Decode} & \textbf{Init} \\
\midrule
Pixel 7 Pro & Gemma 4 E2B & CPU & $\sim$400\,tok/s & $\sim$30\,tok/s & $\sim$10\,s \\
Pixel 7 Pro & Qwen3 0.6B  & CPU & $\sim$200\,tok/s & $\sim$35\,tok/s & $\sim$5\,s \\
\bottomrule
\end{tabular}
\end{table}

These decode speeds mean generating a $\sim$50-token hint takes approximately 1--2 seconds on a Pixel 7 Pro, acceptable for single-turn hint generation. But the initial design required generating complete puzzles (word + category + difficulty + 5 hints as JSON), which produced $\sim$150--200 tokens per puzzle. At 30--35 tokens per second, a single puzzle took 5--7 seconds. Multiplied by 7 puzzles with retries (average 2.5 attempts per successful puzzle), the total generation time exceeded 2 minutes, which is prohibitively slow for a foreground operation.

\subsection{Data Collection and Analysis}

The primary data source is the project's complete Git history: \textbf{204 commits} over 5 days (April 21--26, 2026). Of these, approximately \textbf{90 commits} are directly related to LLM/AI integration (modifying prompts, parsers, model configuration, generation logic, or the engine lifecycle).
\ifanonymized
  The full repository will be made publicly available upon acceptance.
\else
  The full repository is publicly available.\footnote{\url{https://github.com/woliveiras/palabrita}}
\fi

Each AI-related commit was analyzed and coded for: (a)~what failure or limitation triggered the change, (b)~what strategy was applied, and (c)~whether the strategy resolved the issue or led to further iteration. Commits were then grouped into thematic phases and cross-referenced to identify recurring patterns. The commit messages follow the Conventional Commits specification (\texttt{feat:}, \texttt{fix:}, \texttt{refactor:}, \texttt{chore:}), providing structured metadata that aids classification.

In addition to the commit log, the project's specification documents (\texttt{specs/} directory) provide contemporaneous design rationale. When a specification was written to address a problem (e.g., ``spec for static puzzles and background generation,'' written $<$12 hours after the first LLM implementation), the specification itself serves as evidence of the problem's severity and the reasoning behind the chosen solution.

% ============================================================
\section{Development Timeline}
\label{sec:timeline}
% ============================================================

Before presenting the technical findings, we provide a day-by-day narrative of the development process. This timeline contextualizes the failure taxonomy and mitigation strategies by showing how each discovery led to the next iteration.

\subsection{Day 1 (April 21): Ambition Meets Reality}

The first day began with comprehensive specifications written before any Kotlin code. The spec defined a formal state machine for the LLM engine (\texttt{Uninitialized $\rightarrow$ Initializing $\rightarrow$ Ready/Error}), two model tiers (Gemma 4 E2B for premium devices, Gemma 3 1B for compact), and prompts that asked the LLM to generate complete puzzles as JSON:

\begin{lstlisting}[caption={Original prompt from Day 1 specification (Gemma 4 system prompt).}]
You are a word generator for a guessing game.
Always respond using the provided function.
Never add text outside the function call.
\end{lstlisting}

\begin{lstlisting}[caption={Original prompt from Day 1 specification (Gemma 3 compact prompt).}]
You are a word generator for a game.
Return ONLY valid JSON, no extra text.
Schema: {"word": "string", "category": "string",
         "difficulty": number, "hints": [...]}
Rules:
- The word MUST be a common noun in $language,
  $minLength-$maxLength letters
- No proper nouns, no accents, lowercase only
- 5 progressive hints: from vaguest to most specific
- Hints MUST NOT contain the word
\end{lstlisting}

Note the Gemma 4 prompt's reference to ``the provided function,'' a vestige from a function-calling approach that was never implemented. The model had no function to call. This instruction persisted in the spec because it was copied from a cloud LLM integration pattern, and nobody questioned it. It is an early example of how cloud model assumptions polluted the on-device design.

The scaffolding commit created 9 Gradle modules with stub implementations. The core AI integration followed, producing 724 lines of code: \texttt{LlmResponseParser} (2-strategy parser), \texttt{PromptTemplates} (4 prompt functions), and \texttt{PuzzleValidator} (6 validation rules). The LiteRT-LM engine was initialized with CPU-only backend:

\begin{lstlisting}[language=Java,caption={Initial LiteRT-LM engine initialization (Day 1).}]
val config = EngineConfig(
    modelPath = modelPath,
    backend = Backend.CPU(),
    cacheDir = context.cacheDir.absolutePath)
val engine = Engine(config)
engine.initialize()
\end{lstlisting}

The puzzle generator used 3 retries with escalating strategies: standard prompt, ``your previous response was invalid,'' and a simplified prompt with relaxed constraints. Validation checked 6 rules: only lowercase letters, word length within range, no duplicates, exactly 5 hints, hints do not contain the target word, and non-empty category.

\textbf{Within 12 hours}, the first implementation was running and the first spec for static puzzles was written (commit \texttt{69e43d3}), identifying five critical problems with the synchronous generation approach. Less than a day in, and the original design was already cracking. The spec stated:

\begin{quote}
\emph{``The current approach blocks the onboarding screen for 1--3 minutes, shows misleading progress (stuck at 0\%, then jumping), and generates only 7 puzzles that are consumed quickly.''}
\end{quote}

The response was immediate: 50 static puzzles per language were shipped as JSON assets, and a WorkManager-based background generation system was implemented. The LLM was already being demoted from the critical path, less than a day after its first integration.

Also on Day 1, Qwen3 0.6B was added as a third model option (commit \texttt{ccb7010}): 614\,MB, requiring only 2\,GB RAM, versus Gemma 3 1B's 1\,GB file and 4\,GB RAM requirement. A post-game LLM chat feature was also introduced the same day (commit \texttt{588aa21}), allowing players to converse with the model after completing a puzzle. It would survive until Day 5.

\subsection{Day 2 (April 22): Parsing, Prompts, and Sessions}

Day 2 was the most intensive for AI engineering: the parser was rewritten with three fallback strategies, the prompt system underwent its first complete rewrite, and chat sessions replaced isolated inference calls.

The parser evolution was triggered by observing LLM outputs in the wild. A single commit (\texttt{c547d4b}) introduced three stacked parsing strategies:

\begin{enumerate}
  \item \textbf{Code fence stripping:} The LLM wrapped valid JSON in \verb|```json ... ```| markdown blocks. This was the most common format violation.
  \item \textbf{Regex extraction:} The LLM embedded JSON within natural language (``Here is your word: \{...\}''). A regex \verb|\{[^{}]*(?:\{[^{}]*\}[^{}]*)*\}| extracted the first JSON object.
  \item \textbf{Structural parsing:} The LLM returned valid JSON but with keys in the wrong language (\texttt{"palabra"} instead of \texttt{"word"}, \texttt{"dicas"} instead of \texttt{"hints"}). This parser deserialized to a generic \texttt{JsonObject} and inferred fields by value type rather than key name.
\end{enumerate}

On the same day, the prompt was rewritten to remove the function-calling reference and add explicit format instructions:

\begin{lstlisting}[caption={Prompt rewrite 1 (Day 2): explicit JSON instructions.}]
Always respond with ONLY a JSON object.
No markdown, no code fences, no explanation.
The JSON keys MUST be in English:
  "word", "category", "difficulty", "hints".
The values for word, category, and hints
  must be in the requested language.
\end{lstlisting}

The ``CRITICAL'' instruction was added for word length compliance:

\begin{lstlisting}[caption={Prompt rewrite 2 (Day 2): priority markers for constraints.}]
CRITICAL: The word MUST have exactly
  $minLength to $maxLength letters.
Count the letters carefully.
Words with fewer or more letters will be rejected.
\end{lstlisting}

Also on Day 2, the generation system moved from isolated single-turn calls to chat sessions with accumulated context:

\begin{lstlisting}[language=Java,caption={Chat session with contextual retry (Day 2).}]
engineManager.createChatSession(systemPrompt).use { session ->
    for (i in 0 until count) {
        val puzzle = generateSinglePuzzle(
            session = session,
            lastFailureReason = lastFailureReason,
            // ...
        )
    }
}
\end{lstlisting}

When a puzzle failed validation, the specific rejection reason was included in the next prompt:

\begin{lstlisting}[caption={Contextual retry prompt (Day 2).}]
Your previous response was rejected:
  word has 7 letters but we asked for 5.
The word MUST have 5-5 letters.
Try again with a DIFFERENT word.
\end{lstlisting}

Finally, Gemma 3 1B was removed (commit \texttt{766508e}), having lasted less than 8 hours. The evaluation showed Qwen3 0.6B (half the size, half the RAM) produced comparable quality for the constrained hint-generation task.

\subsection{Day 3 (April 23): The Great Simplification}

Day 3 brought the most transformative single commit (\texttt{79f1013}): the removal of \texttt{category} and \texttt{difficulty} from the puzzle schema, and the reduction from 5 to 3 hints. This reduced the JSON output schema from 5 fields to 2:

\begin{lstlisting}[caption={Schema evolution: Day 1 vs Day 3.}]
// Day 1 (7 fields):
{"word": "...", "category": "...", "difficulty": N,
 "hints": ["h1","h2","h3","h4","h5"]}

// Day 3 (2 fields):
{"word": "...", "hints": ["h1","h2","h3"]}
\end{lstlisting}

The rationale for each removal:

\begin{itemize}
  \item \texttt{category} was defined in the schema but never used in gameplay. It existed because the spec included it ``for future use,'' but it added a field the LLM could get wrong without any benefit.
  \item \texttt{difficulty} as a 1--5 numeric scale was too abstract for SLMs. The model often returned arbitrary values, and the game had already moved to difficulty-by-word-length (4-letter words are easier than 6-letter words), making the field redundant.
  \item \texttt{rarity} (an early prompt parameter) was removed because SLMs cannot reliably distinguish between common and rare words.
  \item Hints 4 and 5 were reliably poor quality. The first three hints were usually good; additional hints were often repetitive, overly specific (effectively giving away the word), or grammatically broken.
\end{itemize}

The validator was also relaxed: from ``exactly 5 hints'' to ``at least 3 hints,'' and the category check was removed entirely.

On the same day, generation was restructured to be level-based with fixed word lengths per cycle:

\begin{lstlisting}[language=Java,caption={Level-based generation (Day 3).}]
val LEVELS = listOf(
    4 to 5,   // Cycle 0: 4-letter words, batch of 5
    5 to 10,  // Cycle 1: 5-letter words, batch of 10
    6 to 10   // Cycle 2+: 6-letter words, batch of 10
)
\end{lstlisting}

Mixed word lengths within a single batch had been unstable, as the model needed to context-switch between different constraints. A single word length per cycle was simpler and more reliable.

\subsection{Day 4 (April 24): The WorkManager Nightmare}

Day 4 was almost entirely consumed by debugging WorkManager integration. Seven consecutive fixes in a single day, each addressing a different manifestation of the same fundamental problem: WorkManager's persistent state model was incompatible with foreground LLM operations. In hindsight, we should have recognized the mismatch sooner. This is detailed in \cref{sec:workmanager}.

Two prompt improvements also landed on Day 4:

\begin{lstlisting}[caption={Prompt rewrite 3 (Day 4): concrete rejection examples.}]
CRITICAL RULE 1: Word length.
The word MUST have $minLength to $maxLength letters.
Count every letter before answering.
A word like "estufa" has 6 letters
  and would be REJECTED.
CRITICAL RULE 2: Hints.
NEVER include the word itself inside any hint.
\end{lstlisting}

\begin{lstlisting}[caption={Prompt rewrite 4 (Day 4): full language names.}]
// Before:
"Output language for values: pt"

// After:
"Output language for values: Brazilian Portuguese"
"The word MUST be a common noun
  in Brazilian Portuguese"
\end{lstlisting}

The language name mapping was extracted into a reusable function:

\begin{lstlisting}[language=Java,caption={Language name resolution (Day 4).}]
private val LANGUAGE_NAMES = mapOf(
    "pt" to "Brazilian Portuguese",
    "en" to "English",
    "es" to "Spanish"
)
\end{lstlisting}

UTF-8 sanitization was also added (commit \texttt{e2809da}), removing U+FFFD replacement characters that the LiteRT-LM tokenizer occasionally produced and that broke the JSON parser.

\subsection{Day 5 (April 25--26): Architecture Cleanup and the Final Pivot}

The last two days brought three major architectural changes:

\textbf{First, WorkManager was removed entirely} (commit \texttt{6ce6435}, $-$767/$+$576 lines). The replacement was a clean \texttt{GeneratePuzzlesUseCase} using Kotlin coroutines. All seven WorkManager bug fixes became irrelevant overnight.

\textbf{Second, the final pivot} (commit \texttt{a9b9f2d}): word selection was moved entirely to curated word lists shipped as JSON assets. The LLM's sole remaining job was writing three short hints about a word it was given. This eliminated word length violations, word repetition, nonexistent word generation, and wrong-language word generation, all in a single architectural change.

The prompt system was completely replaced:

\begin{lstlisting}[caption={Final prompt system (Day 5): hint-only generation.}]
// System prompt:
You write short, clear word-game hints in $lang.
Rules:
- ALL hints MUST be written in $lang.
  Never write hints in English or any other language.
- Do NOT include the word in any hint.
- Each hint should be one short sentence.
- Return ONLY a JSON object:
  {"hints": ["hint1", "hint2", "hint3"]}

// User prompt:
Write 3 hints for this word IN $lang.
Do NOT write hints in English
  unless the language is English.
Word: "$word"
\end{lstlisting}

The parser was also specialized for hint-only responses, searching for hint-like keys in multiple languages (\texttt{hints}, \texttt{dicas}, \texttt{pistas}, \texttt{clues}, \texttt{indices}, \texttt{consejos}) and supporting pipe-delimited and numbered-list formats as additional fallbacks.

\textbf{Third, the chat feature was removed} (commit \texttt{7876704}). Post-game conversation with the LLM had been implemented with real streaming, but with the pivot to hint-only generation, single-turn inference was sufficient. The chat added complexity (streaming, context management, session lifecycle) without clear user value in a word-guessing game.

A \texttt{HintFallbackProvider} was added as the final safety net:

\begin{lstlisting}[language=Java,caption={Deterministic fallback hints (Day 5).}]
override fun fallbackHints(
    word: String, language: String
): List<String> = when (language) {
    "pt" -> listOf(
        "E algo que as pessoas conhecem",
        "Pode ser encontrado no dia a dia",
        "Tem ${word.length} letras")
    "es" -> listOf(...)
    else -> listOf(
        "It is something people know",
        "It can be found in everyday life",
        "It has ${word.length} letters")
}
\end{lstlisting}

With this fallback, the game is fully playable even if the LLM never produces a single successful response. The LLM enhances the experience; it does not gate it.

% ============================================================
\section{Findings: Failure Taxonomy}
\label{sec:rq2-rq3-findings}
% ============================================================

We identified five categories of failures that emerged during the integration of on-device SLMs. Each category is documented with symptoms, root cause, observed frequency, code examples, and commit evidence. \Cref{tab:failure-summary} provides an overview.

\begin{table}[ht]
\centering
\small
\caption{Summary of the five failure categories. Frequency scale: Occasional $<$ Frequent $<$ Very Frequent $<$ Systematic; Intermittent = condition-dependent.}
\label{tab:failure-summary}
\begin{tabularx}{\textwidth}{clXcc}
\toprule
\textbf{ID} & \textbf{Category} & \textbf{Core Symptom} & \textbf{Frequency} & \textbf{Severity} \\
\midrule
F1 & Output Format Violations & JSON wrapped in code fences, translated keys & Frequent & Medium \\
F2 & Constraint Violations & Wrong word length, wrong language, repetition & Very Frequent & Critical \\
F3 & Context Quality Degradation & Output quality drops after 3--5 generations & Occasional & Medium \\
F4 & Latency Incompatibility & 1--3 minute UI blocks during generation & Systematic & Critical \\
F5 & Model Selection Instability & Model quality varies drastically across families & Intermittent & High \\
\bottomrule
\end{tabularx}
\end{table}

\subsection{F1: Output Format Violations}

\textbf{Symptoms:} The LLM wraps valid JSON in markdown code fences (\verb|```json ... ```|); uses JSON keys in the prompt's output language instead of English (\texttt{"palabra"} instead of \texttt{"word"}, \texttt{"dicas"} instead of \texttt{"hints"}); occasionally produces malformed UTF-8 sequences (U+FFFD replacement characters) that cause the JSON parser to reject otherwise valid responses.

\textbf{Root Cause:} SLMs with limited instruction-following capability reproduce patterns from their training data. Markdown code fences are ubiquitous in training corpora containing code. Key translation occurs because the model interprets the output language instruction as applying to \emph{all} output, including JSON keys. UTF-8 issues appear to be a LiteRT-LM tokenizer/decoder edge case where the token boundary falls mid-character.

\textbf{Representative Examples:}

\begin{lstlisting}[caption={F1 Example 1: JSON wrapped in markdown code fences.}]
```json
{"word": "gato", "hints": ["It has four legs",
  "It purrs", "It is a pet"]}
```
\end{lstlisting}

\begin{lstlisting}[caption={F1 Example 2: JSON keys translated to Portuguese.}]
{"palabra": "gato", "dicas": ["Tem quatro patas",
  "Ronrona", "E um animal domestico"]}
\end{lstlisting}

\begin{lstlisting}[caption={F1 Example 3: UTF-8 corruption.}]
{"word": "ga\xEF\xBF\xBDto", "hints": [...]}
\end{lstlisting}

\textbf{Frequency:} Code fences appeared in the majority of early inference calls. After adding ``No markdown, no code fences'' to the prompt, frequency dropped to $\sim$20--30\%. The structural parser (Strategy~3 in \cref{sec:strategies}) was needed for approximately 15--20\% of successful parses with Qwen3 0.6B. UTF-8 corruption was rare ($<$5\%) but fatal without sanitization.

\textbf{Evidence:} Commit \texttt{c547d4b} (three parsing strategies stacked), commit \texttt{e2809da} (UTF-8 sanitization).

\subsection{F2: Constraint Violations}

\textbf{Symptoms:} The LLM generates words with the wrong number of letters (e.g., 7 letters when 5 were requested); generates words in the wrong language; repeats previously generated words within a batch; includes the target word inside its own hints; generates nonexistent words.

\textbf{Root Cause:} SLMs have weak numerical reasoning. They cannot reliably count characters. The word ``estufa'' has 6 letters, but a 600M parameter model may ``count'' it as 5 or 7. Language drift is an emergent property of English-dominant training data. Repetition increases as the generation context grows and earlier outputs saturate the model's attention.

\textbf{Representative Examples:}

\begin{lstlisting}[caption={F2 Example: prompt asks for 5-letter word, model returns 7.}]
// Prompt: "word MUST have exactly 5 letters"
// Response:
{"word": "ventana", "hints": [...]}
// "ventana" has 7 letters, not 5
\end{lstlisting}

\begin{lstlisting}[caption={F2 Example: hint contains the target word.}]
// Word: "gato"
// Hint: "A gato is a common pet"
// VIOLATION: hint contains the word "gato"
\end{lstlisting}

\textbf{Frequency:} Word length violations occurred in 30--50\% of early generations. The model could not count. After prompt hardening (concrete examples, ``CRITICAL'' markers), violations dropped to $\sim$10--15\%, which was still too high for production reliability. This persistent $\sim$10--15\% failure rate was the primary motivation for the final pivot to curated word lists, which reduced word-related violations to 0\%.

\textbf{Evidence:} Commits \texttt{7dfb417} (``CRITICAL'' instruction added), \texttt{f846666} (concrete rejection examples), \texttt{0a9bbeb} (full language names), \texttt{79f1013} (category/difficulty removed to reduce violation surface).

\subsection{F3: Context Quality Degradation}

\textbf{Symptoms:} After 3--5 generations within the same conversation session, outputs become repetitive, hints lose relevance, and the model increasingly echoes previous outputs rather than generating novel content. In extreme cases, the model begins producing near-identical responses regardless of the input word.

\textbf{Root Cause:} On-device models operate with limited KV cache (typically 4096 tokens for quantized models on mobile). As the conversation history grows, older context is compressed or truncated, and the model's attention distribution degrades. The effective ``memory'' of the model shrinks as new tokens push out older context. This effect is more pronounced in smaller models (Qwen3 0.6B, with fewer attention heads and less representational capacity) than in larger ones (Gemma 4 E2B).

\textbf{Frequency:} Observed qualitatively across manual testing sessions during development ($N \approx 15$ sessions). Degradation consistently appeared after $\sim$3 generations in Qwen3 0.6B sessions and after $\sim$5--7 generations in Gemma 4 E2B sessions. These are development observations, not controlled measurements. The degradation was gradual, not sudden: hint quality decreased incrementally before becoming obviously repetitive.

\textbf{Evidence:} Commit \texttt{1215899} (session rotation every 3 puzzles), commit \texttt{aa7f2dd} (adjusted to 5).

\subsection{F4: Latency Incompatibility}

The initial approach (generating 7 puzzles synchronously during onboarding) blocked the UI for 1--3 minutes. Progress indicators lied: stuck at 0\%, then jumping to completion. On lower-tier devices, waits were even longer. An onboarding flow that should take 10--15 seconds became a multi-minute dead screen.

The arithmetic was unforgiving. A single puzzle generation (word + 5 hints as JSON) took 15--30 seconds on a Pixel 7 Pro with Qwen3 0.6B on CPU. Multiply by 7 puzzles with an average of 2.5 retries per success, and total time exceeded 2 minutes. No user waits 2 minutes for a word game to start.

Unlike the other failure categories, which affected a percentage of generations, latency affected everyone. Every single user. The fix came in two waves: static puzzles plus WorkManager as an immediate band-aid (commit \texttt{69e43d3}, spec written $<$12 hours after first implementation), then the final pivot to curated word lists plus hint-only generation, which reduced per-puzzle time from 15--30 seconds to 1--2 seconds.

\subsection{F5: Model Selection Instability}

Not all SLMs are equal, even at similar parameter counts. We listed six models in the registry but only two survived testing. The compact model (Gemma 3 1B) lasted less than 8 hours before being replaced by Qwen3 0.6B (half the size, comparable quality for our constrained task). A prompt optimized for Qwen3 might produce worse results with Gemma, and vice versa, because each model's instruction-following behavior is an emergent property of its training data.

The practical consequence: each model in the registry meant a different prompt tuning profile, a different set of format violations, and a different reliability characteristic. Maintaining 6 models was equivalent to maintaining 6 separate AI integrations. We cut to 2 (commits \texttt{766508e}, \texttt{0e56401}).

% ============================================================
\section{Findings: The Progressive Simplification Arc}
\label{sec:rq1-simplification}
% ============================================================

A clear pattern emerged across the 5-day development period: \emph{progressive responsibility reduction} applied to the LLM. This pattern directly answers RQ1: on-device SLMs \emph{are} viable, but only after the developer discovers, through iteration, the minimal scope of work the model can handle reliably.

\begin{table}[ht]
\centering
\caption{Progressive reduction of LLM responsibility over the development period.}
\label{tab:simplification}
\begin{tabular}{clccc}
\toprule
\textbf{Day} & \textbf{LLM Responsibility} & \textbf{Fields} & \textbf{Validation Rules} \\
\midrule
1 & Generate everything & 7 & 6 \\
  & (word, category, difficulty, rarity, hints[5]) & & \\
2 & Generate core puzzle & 5 & Relaxed \\
  & (word, category, difficulty, hints[3]) & & \\
3 & Generate word + hints & 2 & Minimal \\
  & (word, hints[3]) & & \\
5 & Generate hints only & 0 (word gen) & Hint quality \\
  & (word from curated list) & & only \\
\bottomrule
\end{tabular}
\end{table}

Each reduction was triggered by observed failures:

\textbf{Day 1 $\rightarrow$ Day 2:} \texttt{category} was never used in gameplay; \texttt{difficulty} as a 1--5 scale was too abstract for SLMs to interpret consistently; reducing from 5 to 3 hints eliminated the low-quality tail (hints 4 and 5 were reliably poor). The validator was relaxed from ``exactly 5 hints'' to ``2--5 hints'' to avoid rejecting responses with acceptable hint counts.

\textbf{Day 2 $\rightarrow$ Day 3:} Removing \texttt{category} and \texttt{difficulty} cut the JSON schema from 5 to 2 fields, dramatically reducing format violations (F1) and validation failures. The probability of at least one field being malformed drops with each field removed: for $n$ fields with independent format violation probability $p$, the expected correct-response probability is $(1-p)^n$. Reducing from 5 to 2 fields approximately doubled the success rate.

\textbf{Day 3 $\rightarrow$ Day 5:} The final pivot. Word selection was moved entirely to curated word lists shipped as JSON assets. The LLM's sole remaining job: write 3 short hints about a word it is \emph{given}. This eliminated word length violations (F2), word repetition, nonexistent word generation, and wrong-language word generation, all in a single architectural change.

This arc also included feature \emph{removal}, summarized in \Cref{tab:features}:

\begin{table}[ht]
\centering
\caption{Features added and subsequently removed during development.}
\label{tab:features}
\begin{tabular}{lccl}
\toprule
\textbf{Feature} & \textbf{Added} & \textbf{Removed} & \textbf{Reason} \\
\midrule
Post-game LLM chat & Day 1 & Day 5 & Complexity without clear user value \\
LLM word generation & Day 1 & Day 5 & Unreliable at SLM scale \\
Gemma 3 1B support  & Day 1 & Day 1 & Qwen3 0.6B: smaller, comparable quality \\
6-model registry    & Gradual & Day 4 & Only 2 models reliably tested \\
WorkManager         & Day 1 & Day 5 & 7+ bugs from persistent state model \\
\bottomrule
\end{tabular}
\end{table}

\textbf{Answer to RQ1:} Yes, it is viable to use on-device LLMs for a fully offline, private user experience, but viability requires \emph{scoping the LLM's task to what it does reliably}. In our case, the equilibrium point was: ``write three short creative hints about a known word in a specified language.'' This is a bounded creative task with soft quality requirements and a deterministic fallback. The initial ambition (structured multi-field generation with hard constraints) was not viable at the SLM scale.

\textbf{Answer to RQ2:} Five failure categories emerge when running LLM inference locally on Android: (F1)~output format violations, where the model wraps valid JSON in markdown or uses non-English keys; (F2)~constraint violations, where the model ignores length, language, or uniqueness requirements; (F3)~context quality degradation, where output quality drops after repeated generations within a session; (F4)~latency incompatibility, where synchronous generation blocks the UI for minutes; and (F5)~model selection instability, where behaviour and reliability vary drastically across model families and parameter counts.

% ============================================================
\section{Mitigation Strategies (RQ3)}
\label{sec:strategies}
% ============================================================

For each failure category, we developed and iteratively refined specific strategies. All strategies are documented with their corresponding commit evidence and code examples.

\subsection{S1: Multi-Layer Defensive Parsing (F1)}

The final parsing pipeline applies five sequential strategies, each catching cases the previous one missed:

\begin{enumerate}
  \item \textbf{UTF-8 sanitization:} Remove U+FFFD replacement characters that break JSON parsers.
  \item \textbf{Code fence stripping:} Detect and remove \verb|```json ... ```| wrappers, including single-backtick variants.
  \item \textbf{Direct JSON decode:} Standard \texttt{kotlinx.serialization} deserialization to the expected data class.
  \item \textbf{Regex extraction:} Extract the first JSON-like substring from surrounding text using \verb|\{[^{}]*(?:\{[^{}]*\}[^{}]*)*\}|.
  \item \textbf{Structural parsing:} Deserialize as a generic \texttt{JsonObject} and infer field roles by \emph{value types} rather than key names.
\end{enumerate}

The structural parser (strategy~5) was the critical breakthrough. It was designed after observing that Qwen3 0.6B frequently returned keys in the prompt's output language (\texttt{"palabra"}, \texttt{"dicas"}, \texttt{"pistas"}) rather than in English. By ignoring key names entirely and inferring fields by their structural properties, the parser became language-agnostic:

\begin{lstlisting}[language=Java,caption={Structural parser: infer fields by value type.}]
private fun tryStructuralParse(
    text: String
): PuzzleResponse? {
    val obj = json.decodeFromString<JsonObject>(
        text.trim())
    var word: String? = null
    var category: String? = null
    var difficulty: Int? = null
    var hints: List<String>? = null

    for ((_, value) in obj) {
        when (value) {
            is JsonArray -> hints = value.jsonArray
                .mapNotNull { it.jsonPrimitive.contentOrNull }
            is JsonPrimitive -> {
                val intVal = value.intOrNull
                val strVal = value.contentOrNull
                when {
                    intVal != null && difficulty == null ->
                        difficulty = intVal
                    strVal != null && word == null
                        && strVal.length in 2..9
                        && !strVal.contains(' ') ->
                        word = strVal
                    strVal != null && category == null ->
                        category = strVal
                }
            }
        }
    }
    if (word != null && hints != null
        && hints.size >= 2)
        return PuzzleResponse(word, hints)
    return null
}
\end{lstlisting}

The inference rules are: an array of strings is always hints; a short string without spaces is the word; an integer is the difficulty; any remaining string is the category. These heuristics are simple but remarkably effective because the puzzle schema has structurally distinct field types.

After the pivot to hint-only generation, the parser was further specialized. It searches for hint-like keys in multiple languages and also parses non-JSON formats:

\begin{lstlisting}[language=Java,caption={Hint parser: multi-language key search and alternative formats.}]
val hintsKeys = setOf("hints", "dicas", "pistas",
    "clues", "indices", "consejos")

// Also parses pipe-delimited format:
// "hint1 | hint2 | hint3"
// And numbered-list format:
// "1. hint1\n2. hint2\n3. hint3"
\end{lstlisting}

The evolution of the parsing pipeline illustrates a broader principle: \textbf{parsing for SLMs must be designed as a defensive pipeline, not a schema validator.} Each layer catches a specific failure mode that the previous layers miss.

\subsection{S2: Contextual Retry with Failure Feedback (F2)}

Instead of retrying with the same prompt on validation failure, the system includes the specific rejection reason in the retry prompt. Within a chat session, the model retains context from previous turns, so the rejection feedback is additive:

\begin{lstlisting}[caption={Contextual retry with failure reason.}]
Your previous response was rejected:
  word has 7 letters but we asked for 5.
The word MUST have exactly 5 letters.
Try again with a DIFFERENT word.
\end{lstlisting}

We observed significantly higher compliance on retry compared to blind repetition, particularly for word length constraints. The model ``understood'' (within the limits of next-token prediction) that it needed to produce a shorter word.

The system also maintains a \texttt{localRejectedWords} set within each generation batch:

\begin{lstlisting}[language=Java,caption={Local rejected words tracking.}]
val localRejectedWords = mutableSetOf<String>()
// On validation failure:
localRejectedWords.add(normalizedWord)
// Passed to prompt builder:
val exclusionList = usedWords + localRejectedWords
\end{lstlisting}

This prevents the model from regenerating the same rejected word, which was a common behavior: when told ``your word was rejected,'' the model would often produce the \emph{same} word again if not explicitly told to avoid it.

\subsection{S3: Progressive Prompt Hardening (F2)}

The prompt system evolved through four complete rewrites over 5 days, each responding to a specific observed failure. \Cref{tab:prompt-evolution} traces this evolution.

\begin{table}[ht]
\centering
\small
\caption{The four prompt rewrites and their triggers.}
\label{tab:prompt-evolution}
\begin{tabularx}{\textwidth}{clX}
\toprule
\textbf{Day} & \textbf{Failure Observed} & \textbf{Prompt Change} \\
\midrule
1 & Model doesn't know what function to call & Remove function-calling reference; add explicit JSON schema \\
2 & Code fences, wrong key names & ``No markdown, no code fences''; ``Keys MUST be in English'' \\
2 & Wrong word length & ``CRITICAL: Count the letters carefully'' \\
3 & Still wrong length & Concrete example: ``estufa has 6 letters and would be REJECTED'' \\
4 & Output in English when Portuguese requested & Change ``pt'' to ``Brazilian Portuguese'' in all positions \\
4 & Language drift in hints & Repeat language requirement in both system and user prompt \\
5 & N/A (pivot) & Entire prompt replaced for hint-only generation \\
\bottomrule
\end{tabularx}
\end{table}

Key findings about prompt engineering for SLMs:

\begin{itemize}
  \item \textbf{Priority markers matter.} The word ``CRITICAL'' before a constraint measurably improved compliance, suggesting SLMs are sensitive to emphasis tokens. This is consistent with the hypothesis that these tokens receive high attention weights due to their rarity and association with important instructions in training data.
  \item \textbf{Concrete examples beat abstract rules.} ``Count the letters carefully'' was less effective than ``A word like `estufa' has 6 letters and would be REJECTED.'' The concrete example provides a worked example that the model can pattern-match against, rather than requiring the model to operationalize an abstract instruction.
  \item \textbf{Full language names beat ISO codes.} SLMs interpret \texttt{"pt"} ambiguously (Portuguese? Part? Point?). \texttt{"Brazilian Portuguese"} is unambiguous and maps to a clear language cluster in the model's training data.
  \item \textbf{Repetition across prompt sections reinforces.} Specifying the target language in \emph{both} the system prompt and the user prompt reduced language drift significantly compared to specifying it only once. For the final hint-only prompt, the language appears three times: in the system prompt rules, in the user prompt instruction, and in a negative instruction (``Do NOT write hints in English unless the language is English'').
\end{itemize}

\subsection{S4: Session Rotation (F3)}

To combat context degradation, the system creates fresh conversation sessions at fixed intervals:

\begin{lstlisting}[language=Java,caption={Session rotation implementation.}]
while (puzzleIndex < count) {
    val chunkSize = remaining
        .coerceAtMost(SESSION_ROTATION)
    engineManager.createChatSession(systemPrompt)
        .use { session ->
            repeat(chunkSize) {
                generateSinglePuzzle(session, ...)
                puzzleIndex++
            }
        }
}

companion object {
    private const val SESSION_ROTATION = 5
}
\end{lstlisting}

The rotation interval was tuned empirically: initially 3 (commit \texttt{1215899}), later adjusted to 5 (commit \texttt{aa7f2dd}). The adjustment was motivated by practical batch sizes: with batches of 5 and 10, a rotation of 3 created sessions with only 1--2 puzzles (the remainder of 5/3 and 10/3), wasting the session setup overhead. A rotation of 5 aligned cleanly with both batch sizes.

The tradeoff is that fresh sessions lose the contextual retry benefit (the model no longer ``remembers'' previous rejections). In practice, this tradeoff favored freshness: the quality improvement from a clean KV cache outweighed the loss of retry context for batches larger than 5.

\subsection{S5: Responsibility Reduction (F2, F4)}

The most impactful strategy was \emph{reducing what we asked the LLM to do}. This manifested in two forms:

\textbf{Schema simplification:} Reducing the JSON output schema from 7 fields to 2 (then to hint-only) eliminated entire categories of validation failures. Fewer fields mean fewer opportunities for format violations, constraint violations, and cross-field inconsistencies. The probability argument is straightforward: if each field has an independent $\sim$85\% chance of being correct, a 7-field schema has a $0.85^7 \approx 32\%$ chance of being entirely correct, while a 2-field schema has $0.85^2 \approx 72\%$.

\textbf{Task boundary shift:} Moving word selection from the LLM to curated word lists was the single most effective reliability improvement. The LLM's job changed from ``generate a valid word AND write hints'' (creative + constrained) to ``write hints about THIS word'' (creative only). This separated the deterministic task (selecting a valid word of the right length in the right language) from the creative task (writing natural language hints), assigning each to the appropriate system.

The curated word lists are shipped as JSON assets per language (\texttt{words\_pt.json}, \texttt{words\_en.json}, \texttt{words\_es.json}), validated by a CI script that checks for duplicates, minimum word counts per length, and proper encoding. Adding a new language requires only a JSON file and a manifest entry, with zero Kotlin code changes. The Italian dataset was added as proof of this extensibility (commit \texttt{5907193}: 446 words, zero lines of Kotlin).

\subsection{S6: Guaranteed Deterministic Fallback (All)}

The final architecture includes a \texttt{HintFallbackProvider} that supplies generic template hints when the LLM fails entirely. These hints are generated per language and require zero LLM inference. The game is fully playable even if the LLM never produces a single successful response.

We consider this the core design principle for on-device AI: never make the experience dependent on LLM success. The LLM is a \emph{quality enhancer}, not a \emph{gating dependency}. If it works, hints are creative and contextual. If it fails, hints are generic but functional. In a fast-paced word game, the user may not even notice.

\textbf{Answer to RQ3:} Six strategies effectively mitigate the failure categories: (S1)~multi-layer defensive parsing to handle format violations; (S2)~contextual retry with failure feedback to improve constraint compliance; (S3)~progressive prompt hardening to progressively narrow the model's output space; (S4)~session rotation to reset context degradation before it accumulates; (S5)~responsibility reduction to eliminate entire failure classes by shrinking the LLM's task scope; and (S6)~guaranteed deterministic fallback to ensure the feature is always functional regardless of LLM success.

% ============================================================
\section{Case Study: The WorkManager Anti-Pattern}
\label{sec:workmanager}
% ============================================================

An instructive side story of this integration is the adoption and subsequent removal of Android's WorkManager for LLM-based puzzle generation. This section documents the pattern because it represents a common architectural mistake when integrating generative AI into mobile applications.

\subsection{Why WorkManager Was Chosen}

When the synchronous generation approach failed (F4), the natural Android solution was WorkManager: Google's recommended API for deferrable, guaranteed background work that survives process death. The reasoning was sound: puzzle generation should happen in the background so the user can play with static puzzles while the LLM generates more.

The implementation used a \texttt{CoroutineWorker} with foreground service support, notifications, progress tracking, and retry logic. It worked correctly for the first few test runs.

\subsection{Seven Bugs in One Day}

Day 4 (April 24) produced seven consecutive bug fixes, each addressing a different manifestation of the fundamental incompatibility between WorkManager's persistent state model and foreground LLM operations:

\begin{enumerate}
  \item \textbf{Progress bar hardcoded to 50:} The progress calculation used a hardcoded denominator of 50 (from the static puzzle count), but actual batches produced 5--10 puzzles. The progress bar never passed 20\%.

  \item \textbf{Stale SUCCEEDED state:} When re-triggering generation, WorkManager returned the \texttt{SUCCEEDED} state from the \emph{previous} execution. The UI showed a false success screen immediately, before any new generation had started.

  \item \textbf{Terminal state in \texttt{init}:} The ViewModel's \texttt{init} block observed the WorkManager state flow and unconditionally set \texttt{isComplete = true} on \texttt{SUCCEEDED}, catching the stale terminal state from a previous run. Fix: added a \texttt{hasTriggered} flag.

  \item \textbf{Engine not initialized after app restart:} After a process death and restart, the LLM engine was in \texttt{Uninitialized} state (it only existed in memory). The WorkManager worker expected a ready engine, causing an infinite retry loop. Fix: initialize the engine before scheduling the worker.

  \item \textbf{Cancellation and re-generation:} The cancellation flow did not properly clean up work requests, leaving orphaned workers that interfered with new generation requests.

  \item \textbf{Double-scheduling:} Without proper guards, the UI could schedule two workers simultaneously. The workers competed for the LLM engine (which supports only one active session), causing one to fail and trigger misleading error states. Additionally, worker output data was lost because the code read from \texttt{progress} instead of \texttt{outputData} for completed workers.

  \item \textbf{Stale terminal state (again):} Even with \texttt{hasTriggered}, the ViewModel processed \texttt{SUCCEEDED} states that preceded the current \texttt{RUNNING} state. Fix: added \texttt{hasSeenRunning} flag, only processing terminal states after at least one \texttt{RUNNING} emission.
\end{enumerate}

\subsection{Root Cause Analysis}

The root cause was a mismatch between WorkManager's design assumptions and the LLM generation use case:

\textbf{WorkManager assumes:} Work is deferrable, survives process death, terminal states (\texttt{SUCCEEDED}, \texttt{FAILED}) are the final answer, and the system queries work status asynchronously.

\textbf{LLM generation requires:} The user watches a progress screen, expects real-time feedback, wants to re-trigger generation, and the engine state (in-memory) does not survive process death.

Each bug fix added a guardrail (flags, state filters, initialization checks), but the guardrails were fighting the abstraction. By Day 5, the worker had accumulated enough defensive code that it was harder to understand than the raw coroutine it was wrapping.

\subsection{Resolution: Use Case with Coroutines}

The replacement (commit \texttt{6ce6435}) was a \texttt{GeneratePuzzlesUseCase} using standard Kotlin coroutines:

\begin{lstlisting}[language=Java,caption={Clean coroutine-based generation (replacing WorkManager).}]
@Singleton
class GeneratePuzzlesUseCaseImpl @Inject constructor(
    private val puzzleRepository: PuzzleRepository,
    private val puzzleGenerator: PuzzleGenerator,
    private val engineManager: LlmEngineManager,
    private val appPreferences: AppPreferences,
) : GeneratePuzzlesUseCase {

    override suspend fun execute(
        language: String,
        modelId: ModelId,
        onProgress: (Int, Int) -> Unit,
    ): GenerationResult {
        val unplayed = puzzleRepository
            .countAllUnplayed(language)
        if (unplayed >= GameRules.REPLENISHMENT_THRESHOLD)
            return GenerationResult(0, -1)

        require(engineManager.isReady())
        // ... generation, retry, persistence ...
    }
}
\end{lstlisting}

The ViewModel became trivial:

\begin{lstlisting}[language=Java,caption={ViewModel after WorkManager removal.}]
// Before (WorkManager): schedule -> observe WorkInfo
//   flow -> filter stale states -> extract progress
//   -> handle terminal states with flags
// After (coroutine):
useCase.execute(language, modelId) { progress, total ->
    _state.update { it.copy(
        progress = progress, total = total) }
}
\end{lstlisting}

All seven WorkManager bug fixes became irrelevant. Gone. No persistent terminal states, no stale state observation, no initialization race conditions. The change removed 767 lines and added 576, with a net reduction of 191 lines and a dramatic reduction in complexity.

\subsection{Lesson}

\textbf{WorkManager is not for foreground operations.} If the user is watching a progress screen, the operation is foreground, full stop, even if it involves an LLM running ``in the background.'' The persistence guarantees that make WorkManager valuable for truly deferrable tasks (uploading analytics, syncing data) become liabilities for interactive operations. Standard Kotlin coroutines with structured concurrency are the correct tool for this use case.

% ============================================================
\section{Discussion}
\label{sec:discussion}
% ============================================================

\subsection{Design Heuristics for On-Device SLM Integration}

From our findings, we distill eight actionable design heuristics:

\textbf{H1: Minimize LLM output schema complexity.} Each additional field in a structured output schema multiplies the probability of at least one field being malformed. For SLMs, aim for the smallest schema that serves the feature. In our case, the optimal schema was a single array of three strings: \texttt{\{"hints": ["h1", "h2", "h3"]\}}.

\textbf{H2: Stack multiple parsing fallbacks.} Never assume format compliance from an SLM. Design a parsing pipeline where each layer catches failures the previous one missed. Our pipeline evolved from 1 strategy to 5 over the development period. The structural parsing approach (inferring fields by value type rather than key name) is particularly effective for multilingual scenarios where the model may translate JSON keys.

\textbf{H3: Use contextual feedback in retries.} When a generation fails validation, include the specific failure reason in the next prompt. With conversation sessions, the model retains this context and adjusts. This is dramatically more effective than blind retry, where the model has no information about what went wrong and may repeat the same error.

\textbf{H4: Rotate inference sessions to prevent degradation.} On-device models with limited KV cache degrade after several turns. Creating fresh sessions at fixed intervals (3--5 generations in our experience) maintains output quality. The optimal interval depends on the model's context window and the average response length.

\textbf{H5: Use full language names, concrete examples, and priority markers.} SLMs respond better to ``Brazilian Portuguese'' than ``pt'', to ``the word `estufa' has 6 letters and would be REJECTED'' than ``count letters carefully'', and to ``CRITICAL:'' before important constraints. These techniques exploit the model's sensitivity to specific tokens and its ability to pattern-match from concrete examples.

\textbf{H6: Always provide a deterministic fallback.} The core user experience should never depend on LLM success. Design a fallback that requires zero inference and keeps the feature functional. In our case, generic template hints keep the game playable even with 100\% LLM failure.

\textbf{H7: Let the LLM do creative work; use deterministic sources for factual work.} SLMs are unreliable for constrained factual tasks (selecting a valid 5-letter word in Portuguese) but adequate for bounded creative tasks (writing three hints about a given word). Identify the creative vs. deterministic boundary in your feature and assign each to the appropriate system.

\textbf{H8: Fewer supported models means a smaller bug surface.} Each model has unique instruction-following behavior and requires independent prompt tuning. We registered 7 models, tested 3, and shipped 2. Each additional model multiplied the testing and prompt engineering burden. Support the minimum viable set of models that covers your device range.

\subsection{Answering RQ1: Viability Assessment}

On-device SLMs are viable for production mobile applications today, with significant caveats:

\textbf{Viable for:} Bounded creative text generation with soft quality requirements and deterministic fallback. Our hint generation feature (write three short hints about a known word in a specified language) represents this sweet spot. The quality bar is ``helpful but not perfect'' rather than ``must be exactly correct.''

\textbf{Not viable for (at current SLM scale):} Multi-field structured generation with hard constraints (exact word length, precise language compliance, no repetition). Our initial design required the LLM to simultaneously satisfy lexical, numerical, linguistic, and structural constraints, and this exceeded the reliable capability of models at the 0.6B--2.6B parameter range.

\textbf{Viability conditions:} (a)~The developer must be prepared for extensive prompt iteration (four complete rewrites in our case); (b)~parsing must be designed as a defensive pipeline, not a single deserializer; (c)~the feature must degrade gracefully when the LLM fails; (d)~the LLM's scope of responsibility must be discovered empirically through iteration, not assumed from cloud model experience.

\subsection{Framework Choice as an Enabler}

The decision to use LiteRT-LM (\cref{sec:framework-eval}) proved critical for two reasons:

First, \textbf{full prompt control} enabled the four prompt rewrites documented in \cref{sec:strategies}. Although AICore's ML Kit now offers a Prompt API (beta) for free-form prompts, it exposes no sampler configuration (temperature, topK, topP), which would have prevented the fine-grained inference control essential for this work. The ability to adjust system prompts, user prompts, temperature, topK, and topP independently was essential. In a development cycle where prompts were rewritten every 1--2 days, any friction in the prompt-to-inference pipeline would have been a bottleneck.

Second, \textbf{model flexibility} enabled the rapid Gemma 3 $\rightarrow$ Qwen3 swap (Section~\ref{sec:models}, F5). This swap required changing only the model configuration (no framework migration, no recompilation pipeline). With MLC LLM, swapping models requires re-running the TVM compilation pipeline, which adds hours to each model evaluation. With AICore, model choice is not an option at all.

A third benefit emerged during testing: \textbf{emulator support}. LiteRT-LM's CPU backend works on emulators, enabling rapid iteration without a physical device for basic prompt testing. MLC LLM explicitly documents that emulators are unsupported, and AICore requires specific flagship hardware across select vendors. For a solo developer, the ability to test on an emulator during early iteration was a significant productivity advantage.

\subsection{Implications for Framework Developers}

Our experience suggests several improvements that inference framework developers could make:

\begin{itemize}
  \item \textbf{Built-in output format enforcement.} Frameworks could offer JSON-mode or schema-constrained decoding (constraining the model's output vocabulary to valid JSON tokens at each step). This would eliminate F1 entirely at the framework level.
  \item \textbf{Session quality metrics.} Exposing KV cache utilization or a ``session freshness'' metric would help developers implement session rotation without empirical tuning.
  \item \textbf{Backend failure diagnostics.} Our GPU backend worked for initialization but failed during inference on emulators (missing OpenCL libraries). Better pre-flight checks or diagnostic error codes would reduce debugging time.
  \item \textbf{Model compatibility metadata.} A machine-readable format describing each model's known limitations (e.g., ``weak at character counting,'' ``tends to translate JSON keys'') would help developers select models and design prompts.
\end{itemize}

\subsection{Implications for Model Developers}

For SLM developers targeting on-device deployment:

\begin{itemize}
  \item \textbf{Character counting is a training gap.} The inability to count letters in a word is a consistent failure across all models we tested. Fine-tuning on character-counting tasks or including such examples in instruction-tuning data would improve constrained generation.
  \item \textbf{Instruction language should not bleed into output structure.} When told to ``output in Brazilian Portuguese,'' models should apply this to \emph{values} but not to JSON \emph{keys}. This distinction is not well-captured in current instruction-tuning approaches.
  \item \textbf{Format compliance training should include mobile-specific patterns.} Models are trained on web corpora where JSON is almost always wrapped in markdown code fences. Explicit training to \emph{not} produce code fences when instructed would help mobile use cases.
\end{itemize}

\subsection{The Gap Between Vision and Reality}
\label{sec:ambition-gap}

We started this project with genuine excitement. On-device LLMs promised a game where every puzzle was unique, every hint was contextual, and the AI could even chat with the player after a round. No cloud dependency, no API costs, full privacy. We gave ourselves five days, a realistic one-week sprint, because we wanted this study to reflect the conditions a working software engineer actually faces: finite time, competing priorities, and the pressure to ship.

\Cref{tab:ambition-gap} shows the gap between what we envisioned on Day~0 and what we actually shipped on Day~5.

\begin{table}[ht]
\centering
\small
\caption{Desired features vs.\ shipped features after the 5-day sprint.}
\label{tab:ambition-gap}
\begin{tabularx}{\textwidth}{lccX}
\toprule
\textbf{Feature} & \textbf{Desired} & \textbf{Shipped} & \textbf{Why / What Would Be Needed} \\
\midrule
LLM generates words & Yes & No & 30--50\% word-length violation rate (F2). Would need a model with reliable character counting, or constrained decoding at the framework level. \\
\addlinespace
LLM assigns category & Yes & No & Field added validation surface without gameplay value. Could work if category were a soft label (no validation), but we had no UI for it. \\
\addlinespace
LLM assigns difficulty (1--5) & Yes & No & SLMs returned arbitrary values. Would need fine-tuning on difficulty calibration data or a post-hoc heuristic (word frequency as proxy). \\
\addlinespace
5 progressive hints & Yes & 3 & Hints 4 and 5 were consistently low quality (repetitive, too specific, or grammatically broken). A larger model ($\geq$7B) or retrieval-augmented hints might sustain 5. \\
\addlinespace
Post-game LLM chat & Yes & No & Worked technically (streaming, multi-turn), but added complexity (session lifecycle, context management) without clear user value in a word game. Would need a compelling game mechanic to justify the cost. \\
\addlinespace
6 model options & Yes & 2 & Each model required independent prompt tuning and testing. With more time (2--3 sprints), we could have qualified 4--5 models with per-model prompt profiles. \\
\addlinespace
Background generation via WorkManager & Yes & No & 7 bugs in one day from persistent state mismatch (Section~\ref{sec:workmanager}). Would need a custom foreground service or a redesigned WorkManager contract with explicit session semantics. \\
\addlinespace
Fully LLM-driven game & Yes & Hybrid & The LLM enhances (hints) but does not drive (word selection). A viable fully-driven game would need models at $\geq$7B parameters with constrained decoding support, or a server-side validation layer (which defeats the offline goal). \\
\bottomrule
\end{tabularx}
\end{table}

Of the eight features we envisioned, we shipped three in reduced form (3 hints instead of 5, 2 models instead of 6, hybrid instead of fully LLM-driven) and cut five entirely. That is a 62\% feature cut rate. This is not a failure of planning; it is the natural outcome of honest contact with a technology whose production characteristics are poorly documented. We planned ambitiously because the demos and benchmarks suggested it was possible. We shipped pragmatically because the code told us otherwise.

The five-day constraint was deliberate. We did not want an open-ended research timeline; we wanted to experience what a developer faces when a product manager says ``ship it by Friday.'' Every cut in \cref{tab:ambition-gap} was made under that pressure. A longer timeline would not have eliminated the failures, only delayed the reckoning. The core lesson, that SLMs require radical scope reduction, would have emerged regardless.

\subsection{Limitations of the Progressive Simplification Pattern}

The progressive simplification arc (\cref{tab:simplification}) raises a question: could we have started at the minimal scope and avoided the 4-day iteration? In retrospect, partially. A developer applying heuristics H1, H6, and H7 from the start would begin with curated word lists and hint-only generation, potentially compressing the iteration cycle from 5 days to 1--2 days.

Still, the \emph{process} of discovering that scope was informative, and we would not trade it. The initial ambitious design revealed the specific failure modes and their severity, which informed the heuristics. Without attempting full puzzle generation, we would not have quantified the 30--50\% word-length violation rate, would not have discovered the KV cache degradation pattern, and would not have developed the structural parsing technique. The heuristics are, in a sense, the compressed experience of the full iteration.

\subsection{Prompt Engineering as Empirical Science}

Our prompt evolution (\cref{tab:prompt-evolution}) confirms that prompt engineering for SLMs is an empirical process, not a design-time activity. Each rewrite was a hypothesis (``if I add CRITICAL, the model will comply with length constraints'') tested against real model outputs. Some hypotheses held up (priority markers improved compliance), others only partially (concrete examples helped but did not eliminate violations), and the overall direction was always toward simplification: fewer instructions, simpler tasks.

This means prompt engineering time should be budgeted into any SLM integration project. In our case, prompt-related work consumed roughly 30--40\% of the AI-related development effort. For cloud models, prompts are often write-once; for SLMs, they are an iterative, ongoing concern.

% ============================================================
\section{Threats to Validity}
\label{sec:threats}
% ============================================================

\textbf{Construct validity.} The study analyzes a single project by a single developer, which limits the diversity of perspectives and approaches. However, the commit log provides a contemporaneous, objective record of decisions and changes that reduces retrospective bias. The Conventional Commits format and Spec Driven Development workflow provide additional structure that aids analysis.

\textbf{External validity.} Findings are derived from specific models (Gemma 4 E2B, Qwen3 0.6B) running on specific hardware (Pixel 7 Pro) via a specific framework (LiteRT-LM). A Pixel 10 Pro was available during development but systematic benchmark collection was not performed on it; benchmark data therefore reflects a single-device configuration. While we expect the failure categories (F1--F5) to generalize across SLMs and frameworks (format violations and constraint violations are inherent to small models, not specific to LiteRT-LM), the specific mitigation strategies may require adaptation for other contexts. The failure frequencies reported are estimates based on development observations, not controlled experiments with statistically significant sample sizes.

\textbf{Internal validity.} As a single-developer project, we cannot disentangle developer skill from strategy effectiveness. A less experienced developer might discover different failure patterns or require different mitigation strategies. Conversely, a team setting might surface failures faster through diverse testing. The 5-day development window means that strategies were evaluated based on rapid iteration rather than long-term monitoring.

\textbf{Reliability.} The 5-day development window is intensive but short. Longer-term maintenance may surface additional failure categories (e.g., model version regressions, hardware-specific inference bugs). We have already observed one such case: GPU backend initialization succeeds on emulators via WebGPU/llvmpipe but inference fails due to missing OpenCL libraries, requiring emulator-specific CPU-only fallback. The model ecosystem is also evolving rapidly; newer SLMs may not exhibit the same failure patterns documented here.

% ============================================================
\section{Conclusion}
\label{sec:conclusion}
% ============================================================

We presented an empirical case study of integrating on-device Small Language Models into a production Android application. Over 204 commits in 5 days ($\sim$90 directly AI-related), we identified five failure categories specific to local SLM inference: output format violations, constraint violations, context quality degradation, latency incompatibility, and model selection instability. We documented the prompt engineering and architectural strategies that mitigate them.

Our central finding is a \emph{progressive simplification arc}: the LLM's responsibility was systematically reduced from structured multi-field generation (7 fields, 6 validation rules) to hint-only generation (0 fields for word generation, hint quality as the only metric) as the developer discovered, through iteration, the boundary of reliable SLM performance. This arc converged on a design principle: \textbf{the most reliable on-device LLM feature is one where the LLM does the least.}

Along the way, we documented a complete prompt engineering iteration cycle (four rewrites in five days), a parsing pipeline that evolved from one strategy to five, a WorkManager integration that produced seven bugs in a single day before being replaced entirely, and a model evaluation that started with nine candidates and converged on two.

On-device SLMs are viable for production mobile applications today. The privacy guarantee is real (no data leaves the device). The offline capability is real (the game works without any network connection). But achieving this viability required accepting that SLMs are not smaller versions of cloud models; they are fundamentally different tools that require different engineering approaches: defensive parsing instead of schema validation, contextual retries instead of blind repetition, responsibility reduction instead of prompt elaboration, and deterministic fallbacks instead of optimistic trust.

We hope our eight heuristics, the detailed failure taxonomy, and the candid account of the development process help practitioners skip the 5-day discovery process we documented and arrive at robust on-device LLM architectures faster.

\subsection{Privacy and Economic Implications}

Beyond the engineering lessons, this work carries a broader implication that we believe deserves explicit attention: \textbf{the current default of routing user data through cloud AI providers is neither economically sustainable nor ethically defensible at scale.}

The economics are stark. As of April 2026, cloud inference pricing ranges from \$0.75--\$5.00 per million input tokens and \$4.50--\$30.00 per million output tokens for flagship models (OpenAI GPT-5.x), with Google's Gemini 3.1 Pro at \$2.00--\$4.00 input and \$12.00--\$18.00 output per million tokens~\cite{openaipricing,geminipricing}. Even budget-tier models (Gemini 2.5 Flash-Lite at \$0.10/\$0.40 per million tokens) accumulate meaningful costs at scale. For a game like Palabrita, generating three hints per puzzle for 100,000 daily active users would cost hundreds of dollars per day in API fees alone. On-device inference costs exactly zero per query after the initial model download. The marginal cost of serving one more user is zero. This difference is not incremental; it is structural.

But the cost argument, while compelling, is secondary to a more fundamental concern: \textbf{user data privacy and the opacity of cloud AI platforms.}

When a user's input is sent to a cloud API, the user loses control over that data. Google's own Gemini API pricing page states plainly: on the free tier, ``Content used to improve our products''~\cite{geminipricing}. Only paying customers receive the assurance that their data is not used for model training. Most users never read these terms. Most developers never surface them. The default is silent data extraction, and the burden of opting out falls on the user (if an opt-out exists at all).

On-device inference eliminates this vector entirely. In Palabrita, the user's gameplay data (words guessed, hints requested, languages chosen) never leaves the device. There is no API call to intercept, no server log to subpoena, no training pipeline to inadvertently include user data. The privacy guarantee is not a policy promise; it is an architectural fact.

We believe the industry must move a significant portion of its inference infrastructure from centralized cloud to the edge: smartphones, personal computers, IoT devices, local servers, self-hosted instances. The hardware is increasingly capable (our Pixel 7 Pro runs a 2.6B parameter model at 30 tokens per second on CPU alone). The frameworks are maturing (LiteRT-LM, MLC LLM, llama.cpp). The models are shrinking (Qwen3 0.6B fits in 614\,MB). What is missing is the engineering culture: a default assumption that user data should stay on the user's device unless there is a clear, consented, and justified reason to send it elsewhere.

This paper documented the engineering cost of that choice. On-device inference is harder than a cloud API call. Parsing is messier, prompts require more iteration, models are less capable, and fallbacks are essential. But these are engineering problems with engineering solutions. The privacy of millions of users is not an engineering problem. It is a design choice, and we argue it should be the default one.

% ============================================================
% Declarations
% ============================================================
\section*{Declarations}

\subsection*{Funding}
This research received no external funding.

\subsection*{Ethical Approval}
Not applicable.

\subsection*{Informed Consent}
Not applicable.

\subsection*{Author Contributions}
William Oliveira: conceptualization, study design, data collection, data analysis, writing (original draft), writing (review and editing).

\subsection*{Data Availability Statement}
\ifanonymized
  The complete Git history of the Palabrita project will be made publicly available upon acceptance.
\else
  The complete Git history of the Palabrita project is publicly available at \url{https://github.com/woliveiras/palabrita}. A preprint of this manuscript is available on arXiv at \url{https://arxiv.org/abs/2604.24636}.
\fi

\subsection*{Conflict of Interest}
The author declares no conflicts of interest.

\subsection*{Acknowledgments}
The author thanks David Chou (Product Manager, Android AI Core, Google) for identifying an inaccuracy in an earlier version of this manuscript regarding the ML Kit Prompt API and its device support, and for permitting this acknowledgment.

% ============================================================
% References
% ============================================================
\bibliographystyle{plain}

\end{document}